\documentclass[10pt,letterpaper]{article}
\usepackage{opex3}

\begin{document}

\title{Soliton interaction mediated by cascaded four wave mixing with dispersive waves}

\author{A.V. Yulin$^1$, R. Driben$^{2,3}$, B.A. Malomed$^2$, and D.V. Skryabin$^{4,5}$}

\address{$^1$1Centro de F\'{i}sica Te\'{o}rica e Computacional,
Universidade
de Lisboa, Ave. Prof. Gama Pinto 2, Lisboa 1649-003, Portugal\\
$^2$Department of Physical Electronics, Faculty of Engineering, Tel
Aviv University, Tel Aviv 69978,  Israel\\$^3$Department of Physics
and CeOPP, University of Paderborn, Warburger Str. 100, D-33098
Paderborn, Germany\\
$^4$Department of
Physics, University of Bath, Bath BA2 7AY, UK\\
$^5$Weierstrass Institute for Applied Analysis and Stochastics, Mohrenstrasse 39, D10117 Berlin, Germany}




\begin{abstract}
We demonstrate that trapping of dispersive waves between two optical
solitons takes place when resonant scattering of the waves on the solitons
leads to nearly perfect reflections. The momentum transfer from the
radiation to solitons results in their mutual attraction and a subsequent
collision. The spectrum of the trapped radiation can either expand or shrink
in the course of the propagation, which is controlled by arranging either
collision or separation of the solitons.
\end{abstract}








Various aspects of dynamics of ultrashort pulses in photonic crystal fibers
(PCFs), such as the formation of solitons, resonant radiation,
supercontinuum generation, optical rogue waves and others  have been
the subject of intense investigations over the past decade \cite{skr,dud}.
One prominent theme of research in this area has been interaction of
dispersive waves with solitons in strongly non-integrable cases, i.e., far
from the limit of the ideal nonlinear Schr\"{o}dinger equation \cite{skr}.
In particular, theoretical and experimental studies reviewed in Ref. \cite%
{skr} have revealed novel phase-matching conditions which result in
generation of new frequency components from the four-wave mixing of solitons and
dispersive waves in the media with significant higher-order dispersions, and
have demonstrated the crucial role of these processes in the expansion of
supercontinuum spectra generated in PCFs \cite{skr,dud}.

Importantly for the present work, the back action of the Cherenkov radiation
\cite{skr,ahmed,skrs} and of the four-wave mixing \cite%
{yulin,pre,shalva,efimov1,efimov2,efimov3} processes on solitons has also
been considered and revealed plausible applications for the control of the
soliton carrier frequencies and group velocities. In this context, a natural
problem to consider is how multiple re-scattering of dispersive waves on two
or several well separated solitons can be used to mediate interactions
between them, and what spectral and temporal-domain effects can be predicted
in this case. While the short-range soliton interaction through the
overlapping soliton tails has been reported in numerous papers, see, e.g. Refs. \cite%
{luan,dribenmalomed,agr} and references therein, the long-range
soliton-soliton interaction via dispersive optical waves or polariton waves
in the material has not yet attracted the same degree of interest.
Nevertheless, some important results on this topic have been already
published --- in particular, the interaction between solitons through the
radiation in the integrable limit \cite{loh}, binding of solitons through the
dispersive waves generated by fourth-order dispersion \cite{buryak}, and interactions between solitons
mediated by acoustic waves \cite{hard}. More recently, it has been noticed
that fission of $N$-solitons in the presence of strong higher-order
dispersion leads at a later stages of the supercontinuum development to
collisions between solitons \cite{podlip,drib0,demir,drib,drib2}. The impact
of a dispersive wave present in the space between colliding solitons on the
random changes of the soliton amplitude and frequency has been studied in
details in Refs. \cite{demir}, formation of bound multi-peak soliton states
through this mechanism has been reported in Ref. \cite{podlip}, and fusion
of several soliton into a single high-amplitude pulse was reported too \cite%
{drib}. From the results presented in Ref. \cite{podlip,drib0,demir,drib},
one can conclude that dispersive waves induce attraction between solitons.
However, the above-mentioned works studied the problem under conditions
typical for the supercontinuum generation, which involves many poorly
controlled factors. Thus, presently there is no clear understanding of the
basic mechanisms driving the interaction between well-separated solitons,
mediated by the dispersive radiation. Investigation of such mechanisms by
means of analytical and numerical methods is a subject of the present work.
We also investigate spectral reshaping of the dispersive radiation
interacting with two solitons, which reveals some interesting opportunities
for spectral control.

As the model we use a normalized form of the generalized NLS equation, which
includes the third-order dispersion (TOD), with respective coefficient $\beta_{3}$,
and the  Raman effect:
\begin{equation}
i\partial _{z}A+{\frac{1}{2}}\partial _{t}^{2}A-i{\frac{\beta _{3}}{6}}%
\partial _{t}^{3}A+ (1-\theta)|A|^2A+A\int_{-\infty }^{+\infty }R(t^{\prime })|A(t-t^{\prime
})|^{2}dt^{\prime }=0,  \label{A}
\end{equation}%
where the response function, with Raman constant $\theta $, delay times $%
\tau _{1}$ and $\tau _{2}$, and step function $\Theta $, is
\[
R=\theta \frac{\tau _{1}^{2}+\tau _{2}^{2}}{\tau
_{1}\tau _{2}^{2}}\Theta (t)\exp \left( -\frac{t}{\tau _{2}}\right) \sin
\left( \frac{t}{\tau _{1}}\right) .
\]
The analysis presented here pertains to all widths of interacting solitons.
The Raman term affects the dynamics only in the case of subpicosecond
pulses, and it is considered only in simulations displayed below in Fig. 3.

\begin{figure}[tbph]
\centering\includegraphics[width=\columnwidth]{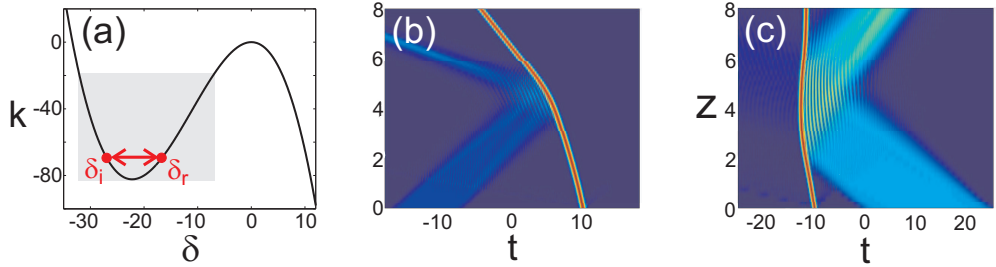}
\caption{ (a) The dispersion diagram and resonances between the reflected
and incident waves predicted by roots of Eq. (\protect\ref{delta}), with $%
\protect\delta =0$ corresponding to the soliton frequency. (b,c) The single
event of the scattering of a dispersive pulse on the soliton for the
incident pulse with frequency $\protect\delta _{r}$ (b) and $\protect\delta %
_{i}$ (c), respectively. Here, $\protect\beta _{3}=-0.015$, the soliton
input is $A=\protect\sqrt{2q}\mathrm{sech}\left( \protect\sqrt{2q}%
(t-t_{s})\right) $, $q=18$, $t_{s}=10$. The input for the dispersive pulse
is $A=\sqrt{I}~\mathrm{sech}\left( (t-t_{0})^{8}/w^{8}\right) e^{i\protect\delta %
_{r,i}t}$, with $t_{0}=12$, and $\sqrt{I}=0.4$, $w=5$ in (b) and $\sqrt{I}=0.5$, $w=12$ in
(c).}
\end{figure}

To understand the interaction of radiation with two solitons, we
first need to recapitulate results for the interaction of a single
dispersive wave packet with a single soliton. The black curve in
Fig. 1 shows the dispersion of linear waves, $A\sim e^{ikx-i\delta t}$, $k(\delta
)=-{(1/2)}\delta ^{2}+\left( \beta _{3}/6\right) \delta^{3}$, where $k$ and $\delta$ are, respectively,
shifts of the propagation constants and frequencies.
Assuming that the soliton has zero frequency detuning, we
have for its wavenumber $k_{s}(\delta )=q$, where $q$ is the soliton
parameter (half the peak power, see the Fig. 1 caption).  For given frequency $%
\delta _{i}$ of the incident wave packet, see Fig. 1(a), one can
derive the phase-matching (resonance) condition for the frequency of
the reflected wave, $\delta _{r}$, as shown in Fig. 1(a) \cite{skr}.
For the resonances due to the the four-wave scattering of the
dispersive wave on the soliton-induced shift of the refractive index
(the only resonances relevant here), we have
\begin{equation}
k(\delta _{r})=k_{s}(\delta _{r})-[k_{s}(\delta_{i})-k(\delta _{i})],
\label{k}
\end{equation}%
two intersection points in Fig. 1 corresponding to roots of this equation.
Taking into regard that the initial soliton momentum is zero and neglecting
the Raman effect, we can  apply the momentum conservation: $%
M_{i}=M_{s}+M_{r}$, where $M=\mathrm{Im}\left\{ \int_{-\infty }^{+\infty
}A\partial _{t}A^{\ast }dt\right\} $, and the weak transmitted wave is
disregarded.  It is also natural to assume that the number of photons
in the soliton, $Q_{s}$, does not change upon the collision. Then, after a
straightforward algebra, the conservation law can be transformed into the
form of $\delta _{i}|k_{i}^{\prime }|IL=\delta _{s}Q_{s}+\delta
_{r}|k_{r}^{\prime }|IL$, where $L$ and $I$ are the spatial lengths and
intensities of the incident and reflected wave packets (which are
approximately equal for both packets), and $k^{\prime }\equiv dk(\delta
)/d\delta $. Transforming this conservation law into a differential form
(for $L\rightarrow 0$), and adopting $|k_{i}^{\prime }|\simeq |k_{r}^{\prime
}|$ (see Fig. 1), we find
\begin{equation}
\partial _{z}\delta _{s}\simeq (\delta _{i}-\delta _{r})I|k_{i}^{\prime
}|/Q_{s}{.}  \label{delta}
\end{equation}%
As follows from Eq. (\ref{delta}), for $\delta _{i}-\delta _{r}>0$ ($\delta
_{i}-\delta _{r}<0$) the soliton's frequency increases (decreases)
resulting, respectively, in the acceleration (deceleration) of the soliton.
Swapping the frequencies of the incident and reflected waves means sending
the incident radiation pulse either ahead of or behind of the soliton, while
its trajectory in the $(t,z)$-plane deviates towards the collision side in
either case, see Figs. 1(b,c).

\begin{figure}[tbph]
\centering\includegraphics[width=\columnwidth]{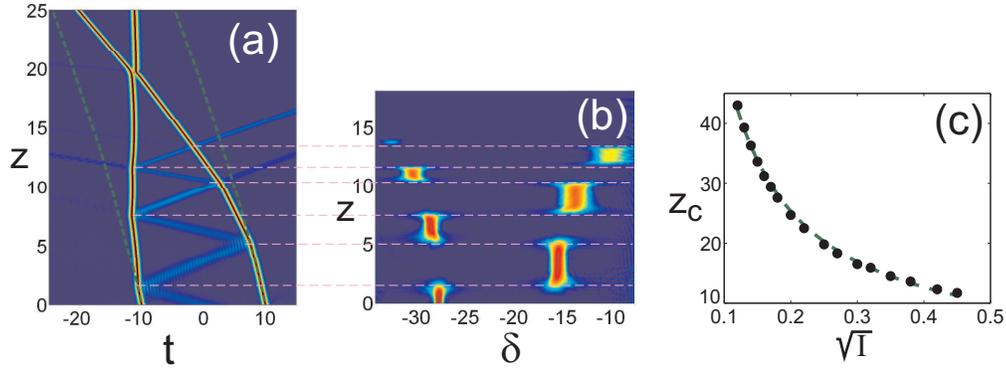}
\caption{ (a) The soliton collision caused by the effective attraction due
to multiple scattering of dispersive waves. Initial solitons have zero
frequencies and $q=12.5$, the dispersive pulse is $A=\sqrt{I}~\mathrm{sech}(t/w)e^{i%
\protect\delta t}$, $\sqrt{I}=0.25$, $\protect\delta =-28$. (b) The change of the
dispersive-pulse's spectrum following the collision shown in (a). (c) The
collision distance ($z_{\mathrm{c}}$) vs. the initial amplitude of the
dispersive pulse ($\sqrt{I}$). For comparison, the best fit to
dependence $z_c\sim 1/\sqrt{I}$ (see the text) is shown in (c) too.}
\end{figure}

Thus, setting two identical well-separated solitons and sending a
pulse between them with any frequency from the shaded interval in
Fig. 1(a),
effective attraction between the solitons is induced by multiple
scattering events between the pulse and the solitons, see Fig. 2(a).
It is important to note that the spectrum of radiation trapped
between the solitons evolves in the course of the propagation, and
can be controlled by the initial choice of the soliton frequencies.
If the two solitons are not identical, then, after two consecutive
scattering events, the radiation comes back to the soliton with an
altered frequency, resulting in spectral evolution of the radiation.
In the case shown in Fig.2, the soliton frequencies drift in the opposite directions
due to interaction with the dispersive waves, which, in
turn, makes the frequency of the dispersive wave varying too, see
Fig. 2(b). Changing the intensity of the dispersive pulse, one can
control the distance at which the solitons eventually collide due to
the attraction, see Fig. 2(c). This control technique may be used if
one needs to produce a high-intensity pulse as a result of the
collision. Since the dispersion law in the soliton frequency
range is quasi-parabolic, the soliton group velocity is proportional to its frequency.
Hence $\partial_z\delta_s$ in the left hand side of Eq.  (3) is in fact
the soliton acceleration or the second derivative of its temporal coordinate and therefore
the effective interaction force $F_{int}$
induced by the radiation pulse and acting between the solitons is proportional to
the intensity of the latter, $I$. Accordingly,
the distance to collision, $z_c$, should naturally scale as $1/\sqrt{F_{int}}\sim 1/\sqrt{I}$, see Fig. 2(c).
We would like to notice here that the effect in question is not sensitive to the exact shape of the
dispersive wave envelope.

\begin{figure}[tbph]
\centering\includegraphics[width=\columnwidth]{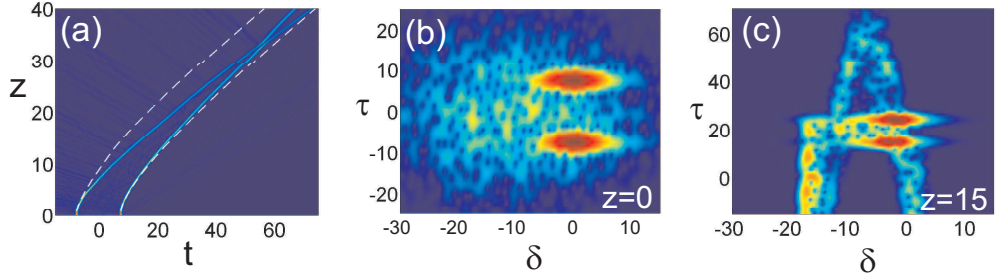}
\caption{The attraction and collision induced by the random wave field
trapped between the solitons. (a) The intensity evolution in the ($z$,$t$%
)-plane for the solitons with zero frequency and $q=8$. (b,c) Initial ($z=0$%
) and intermediate ($z=15$) time-frequency diagrams showing the solitons and
dispersive waves. Density plots in (b,c) show XFROG function calculated as $%
\left\vert \protect\int_{-\infty }^{+\infty }A\exp (-(t-\protect\tau %
)^{2}/w_{r}^{2})\exp (-i\protect\delta t)dt\right\vert $ with $w_{r}=2$. The
parameters are $\protect\beta _{3}=-0.025$, $\protect\theta =0.18$, $\protect%
\tau _{1}=0.061$, and $\protect\tau _{2}=0.16$, cf. Ref. \protect\cite{skr}.
}
\end{figure}

Moreover, in fact it is not necessary to use a dispersive pulse with a
sharply defined frequency and shape to generate the radiation-mediated
attraction between solitons. Instead, one can take a random field with the
spectrum filling a sufficiently broad frequency interval in the range of
normal group-velocity dispersion, see Fig. 1(a). In this case, multiple
waves experience cascaded scattering events with solitons, resulting in
their mutual attraction, as shown in Fig. 3 (in the presence of the Raman
effect). Panel 3(a) shows the evolution in the course of the propagation in
the fiber, while 3(b,c) show XFROG diagrams illustrating spectral-temporal
characteristics of the field at the input and at an intermediate propagation
distance.
Adding the Raman effects (along with the self-steepening) into Eq.
(1) bends the solitons' trajectories but does alter the nature and
outcome of the interaction. This conclusion agrees with the
previously reported studies of the radiation-induced soliton
collisions in the full-scale supercontinuum modeling and experiments
\cite{podlip,drib0,demir,drib}.

\begin{figure}[htbp]
\centering\includegraphics[width=\columnwidth]{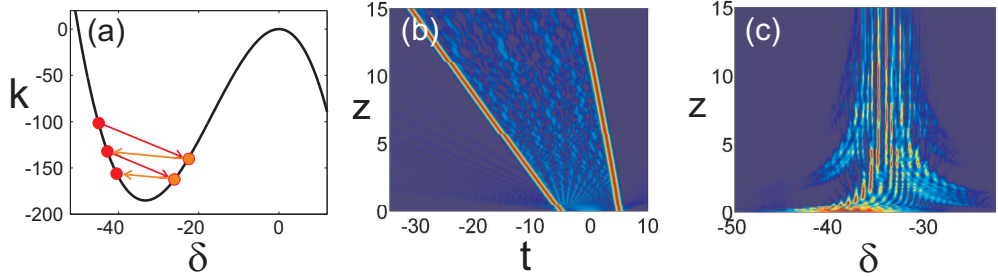} \caption{The
narrowing of the radiation spectrum trapped between two separating
solitons. The initial solitons are taken with $q=12.5$ and
frequencies $\protect\delta=1.5$, $\protect\delta=0$. (a) The
dispersion diagram showing the scattering cascade. (b,c) The
evolution in the temporal and frequency domains. In this case,
$\beta_3=-0.01$.}
\end{figure}

By choosing initial soliton frequencies so that they separate in the
course of the propagation, we have found that, in this case, the
dispersive waves do not strongly affect the motion of the solitons,
see Fig. 4. On the other hand, applying the resonance condition in
panel (a), we find that the frequencies of the dispersive pulse
scattered on the right and left solitons get closer, implying that
the spectrum narrows, as is indeed seen in panel (c).

On the other hand, spectral expansion of the radiation is observed
if the solitons are initially moving towards each other, see Fig. 5.
Panel (a) shows how the resonant scattering modifies the frequency
of the dispersive pulse, and the modification of the spectrum with
the propagation distance is shown in (c).

\begin{figure}[htbp]
\centering\includegraphics[width=\columnwidth]{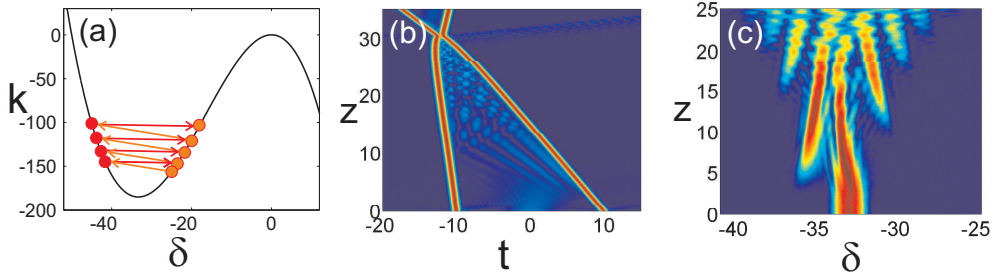} \caption{The same
as in Fig. 4, but for expansion of the radiation spectrum trapped
between two separating solitons. The initial solitons are taken with
$q=18$ and frequencies $\protect\delta=-0.3$, $\protect\delta=0.3$.}
\end{figure}

Summarizing, we have studied effects resulting from the resonant
multiple scattering of dispersive radiation on a pair of well
separated solitons in optical fibers with strong TOD. We have
demonstrated in the numerical form and explained analytically the
effect of the soliton attraction mediated by the multiply scattered
radiation, The spectral reshaping of the radiation trapped between
the two solitons has been investigated too.

DVS acknowledges useful discussions with S. Amiranashvili. The work of AVY
was supported by the FCT grant PTDC/FIS/112624/2009 and
PEst-OE/FIS/UI0618/2011. RD and BAM appreciate funding through grant No.
2010239 from\ the Binational (US-Israel) Science Foundation.


\begin{thebibliography}{99}
\bibitem{skr} D.V. Skryabin and A.V. Gorbach, ``Looking at a
soliton through the prism of optical supercontinuum", Rev. Mod. Phys.
\textbf{82}, 1287-1299 (2010).

\bibitem{dud} J. M. Dudley, G. Genty, and S. Coen, ``Supercontinuum generation
in photonic crystal fiber", Rev. Mod. Phys \textbf{%
78}, 1135-1184 (2006).

\bibitem{ahmed} N. Akhmediev and M. Karlsson, ``Cherenkov
radiation emitted by solitons in optical fibers", Phys. Rev. A \textbf{51},
2602-2607 (1995).

\bibitem{skrs} D.V. Skryabin, F. Luan, J.C. Knight, and P.S. Russell,
``Soliton self-frequency shift cancellation in photonic crystal
fibers", Science \textbf{301}, 1705-1708 (2003).

\bibitem{yulin} A.V. Yulin, D.V. Skryabin, Optics Lett. \textbf{29},
2411-2413 (2004).

\bibitem{pre} D.V. Skryabin and A.V. Yulin, ``Theory of
generation of new frequencies by mixing of solitons and dispersive waves in
optical fibers", Phys. Rev. E \textbf{72}, 016619 (2005).

\bibitem{shalva} A. Demircan, S. Amiranashvili, and G. Steinmeyer,
``Controlling light by light with an optical event horizon", Phys.
Rev. Lett. \textbf{106}, 163901 (2011).

\bibitem{efimov1} A. Efimov, A.J. Taylor, F.G. Omenetto, A.V. Yulin, N.Y.
Joly, F. Biancalana, D.V. Skryabin, J.C. Knight, and P.St.J.
Russell, Optics Express \textbf{12}, 6499 (2004).

\bibitem{efimov2} A. Efimov, A.V. Yulin, D.V. Skryabin, J. C. Knight, N.
Joly, F. G. Omenetto, A. J. Taylor, and P. Russell, Phys. Rev. Lett. \textbf{%
95}, 213902 (2005).

\bibitem{efimov3} A. Efimov and A.J. Taylor, A.V. Yulin, D.V. Skryabin, and
J.C. Knight, Opt. Lett. \textbf{31}, 1624 (2006).

\bibitem{tail-interaction} B. A. Malomed, ``Potential of
interaction between two- and three-dimensional solitons", Phys. Rev. E
\textbf{58}, 7928-7933 (1998).

\bibitem{luan} F. Luan, D.V. Skryabin, A.V. Yulin, and J.C. Knight,
``Energy exchange between colliding solitons in photonic crystal
fibers", Opt. Exp. \textbf{14}, 9844-9853 (2006).

\bibitem{dribenmalomed} R Driben, B. A Malomed ``Suppression
of crosstalk between solitons in a multi-channel split-step system", Opt.
Commun. \textbf{197}, 481-489 (2001)

\bibitem{agr} G. Agrawal, ``Nonlinear Fiber Optics"
(Academic Press, New York, 2007).

\bibitem{loh} W.H. Loh, A.B. Grudinin, V.V. Afanasjev, and D.N. Payne,
``Soliton interaction in the presence of a weak nonsoliton
component", Opt. Lett. \textbf{19}, 698-700 (1994).


\bibitem{buryak} N.N. Akhmediev and A.V. Buryak, ``Interactions of solitons with
oscillating tails", Opt. Commun. \textbf{121},
109-114 (1995).

\bibitem{hard} P.J. Hardman, P.D. Townsend, A.J. Poustie, and K.J. Blow,
``Experimental investigation of resonant enhancement of the acoustic
interaction of optical pulses in an optical fiber", Opt. Lett.
\textbf{21}, 393-395 (1996).

\bibitem{podlip} A. Podlipensky, P. Szarniak, N. Y. Joly, C. G. Poulton, and
P. St. J. Russell, ``Bound soliton pairs in photonic crystal fiber",
Opt. Exp. \textbf{15}, 1653-1662 (2007)

\bibitem{drib0} R. Driben, F. Mitschke, and N. Zhavoronkov,
``Cascaded interactions between Raman induced solitons and
dispersive waves in photonic crystal fibers at the advanced stage of
supercontinuum generation", Opt. Exp. \textbf{18}, 25993-25998
(2010).

\bibitem{demir} A. Demircan, S. Amiranashvili, C. Br\'{e}e, C. Mahnke, F.
Mitschke and G. Steinmeyer, ``Rogue events in the group velocity
horizon," Scientific Reports \textbf{2}, 850 (2012).

\bibitem{drib} R. Driben and I.V. Babushkin, ``Accelerated
rogue waves generated by soliton fusion at the advanced stage of
supercontinuum formation in photonic crystal fibers", Opt. Lett. \textbf{37}%
, 5157-5159 (2012).

\bibitem{drib2} R. Driben and N. Zhavoronkov, ``Supercontinuum spectrum control
in microstructure fibers by initial chirp management," Opt. Exp.
\textbf{18} (16), 16733 (2010).



\end{thebibliography}
\end{document}